# Superstring Partons and Multiple Quantisation

Philip E. Gibbs[1]

## Abstract

I consider an algebraic construction of creation and annihilation operators for superstring and *p*-brane parton models. The result can be interpreted as a realisation of multiple quantisation and suggests a relationship between quantisation and dimension. The most general algebraic form of quantisation may eventually be expressed in the language of category theory.

---

[1] e-mail to philip.gibbs@pobox.com

# Introduction

After twelve years of intensive research in superstring theory there is still a deep mystery surrounding its underlying nature. String theory has been found to take on a number of different forms in background space-times with different dimensions and topologies. Much to the delight of string theorists, many, if not all of these forms may be tied together through dimensional reduction and duality [1]. This has led some to speculate that string theory must have a unique general formulation which is purely algebraic and independent of any space-time background topology.

According to the principle of *p*-brane democracy [2], strings and membranes of various dimensions should be included in the final theory and should all be regarded as equally fundamental. At the same time they may be considered composite. Some theorists have explored the consequences of the idea that strings could be composed of discrete partons [3,4]. This may seem contrary to the spirit of string theory but if interpreted correctly it is not. Interesting conclusions have been drawn from such models of bit-strings embedded in space-times and quantised in a light cone gauge.

But if *p*-brane democracy is to be taken seriously, space-time itself should also be regarded as a membrane like any other and may also be thought of as a composite structure. In that case we must look for an algebraic approach independent of space-time embeddings.

## Algebraic Superstring Theory

In previous work I have defined an infinite dimensional Lie superalgebra for the symmetry of closed strings composed of fermionic partons [5]. The algebra was inspired by the principle of event-symmetric space-time which supposes that the diffeomorphism group of general relativity must be extended to the symmetric group of permutations on space-time events. It was argued that this is necessary in order to allow for space-time topology change.

It was noted that the algebra could also be reformulated as an algebra of string creation and annihilation operators. In this form the base elements of the algebra consist of an ordered sequence of fermion creation and annihilation operators $b_i$ $b^*_i$ linked together by arrows which define an arbitrary permutation. A typical element would look like this:

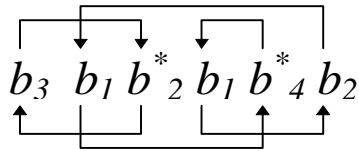

These elements can be multiplied associatively by concatenating them together, but an additional set of relations is enforced which reflect the anti-commutation relations of the creation and annihilation operators. They are defined schematically as follows,

$$b^*_i \, b_j \; + \; b_j \, b^*_i \; = \; 2\delta_{ij}$$

$$b_i \, b_j \; + \; b_j \, b_i \; = \; 0$$

$$b^*_i \, b^*_j \; + \; b^*_j \, b^*_i \; = \; 0$$

Notice that when a creation operator is exchanged with its annihilation partner there is an interaction between the strings. These are partial relations which can be embedded into complete relations. When closed loops which include no operators appear they are identified with unity. For example, the lines of the first equation can be joined to give,

$$b^*_i \, b_j \; + \; b_j \, b^*_i \; = \; 2\delta_{ij}$$

This example shows a cyclic relation on a loop of two operators. The arrows can be joined differently to give another relation,

$$b^*_i \, b_j \; + \; b_j \, b^*_i \; = \; 2\delta_{ij}$$

which is an anti-commutation relation for loops of single operators.

By applying these relations repeatedly it is possible to reorder the operators in any string so that the strings are separated into products and sums of ordered cycles. Therefore we can define a more convenient notation in which an ordered cycle is indicated as follows (where a,b,c represent the creation and annihilation operators),

$$(a\,b\ldots c) \; = \; a \rightarrow b \rightarrow \ldots \rightarrow c$$

We can generate cyclic relations for loops of any length and graded commutation relations between any pair of strings by repeatedly applying the exchange relations for adjacent pairs of operators.

The algebra has a $Z_2$ grading given by the parity of the length of string and it is therefore possible to construct an infinite dimensional Lie-superalgebra using the graded commutator. The algebra may thus be interpreted as both an algebra of creation and annihilation operators and as the supersymmetry algebra of discrete strings.

## Supersymmetry Ladder

The next stage of the algebraic string theory program is to construct a ladder operation which takes us from one supersymmetry algebra to another one. Starting from the one dimensional string supersymmetry constructed in the previous section, the ladder operator will take us up to a symmetry of two dimensional membranes. Further steps take us up to higher dimensional *p*-brane algebras.

We start with a Lie algebra whose elements satisfy the Jacobi relation,

$$[[A,B],C]+[[B,C],A]+[[C,A],B]=0$$

A new algebra is constructed by stringing these elements in a sequence and attaching them with an orientated string passing through each one like before. A difference introduced this time is that the string is allowed to have trivalent branches and we must factor out the following relations,

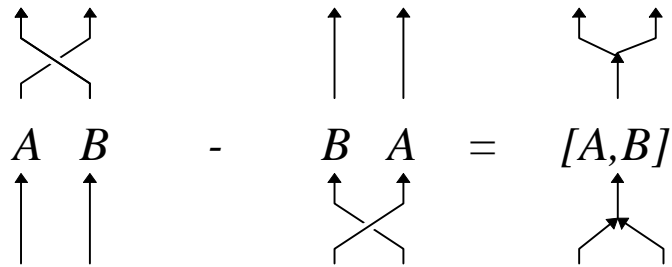

The crossing lines do not (yet) indicated that the lines are knotted. It does not matter which goes over the other.

When we check the result of combining the interchanges of three consecutive elements using the commutation relation above, we find that the result is consistent with the Jacobi relation provided we also apply the following associativity relation

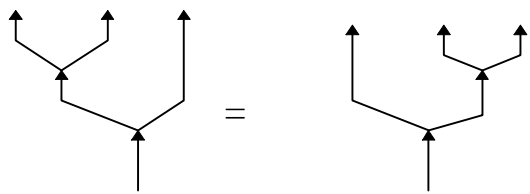

and the co-associativity relation

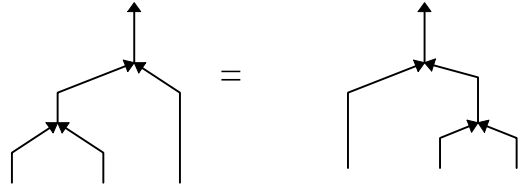

In addition a relationship is used to remove closed loops.

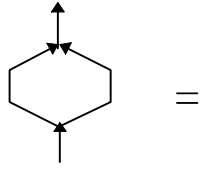

This process defines a new algebra like before except that now we start from any Lie algebra and create a new associative algebra. A new Lie algebra is then defined using the commutators of the algebra as the Lie product.

This construction generalises easily to Lie superalgebras using graded commutators and graded Jacobi relations. Thus we have a ladder operators which maps one superalgebra to a new one. This ladder operator will be signified by $Q$. Thus if $\mathcal{A}$ is a Lie superalgebra then $Q(\mathcal{A})$ is another.

In the case where we start with the discrete string superalgebra the loops can be visualised as circling the new network. These can then be interpreted as sections of a branching string world sheet. The new algebra is therefore a symmetry for string world sheets or membranes. Application of the ladder operator increases the dimension of the structures each time.

The universal enveloping algebra of the original Lie superalgebra is isomorphic to a subalgebra of the higher dimensional one. That is the subalgebra formed by simply looping each element to itself (The loop removing relation is needed to establish this). I.e. there is a mapping from $U(\mathcal{A})$ into $Q(\mathcal{A})$

$$Up: U(\mathcal{A}) \mapsto Q(\mathcal{A})$$

Furthermore, there is a homomorphism from the higher algebra to the universal enveloping algebra below which is defined by removing the string connections.

$$Down: Q(\mathcal{A}) \mapsto U(\mathcal{A})$$

It is possible to apply the ladder operator any number $n$ times giving the algebra $Q^n(\mathcal{A})$. Since the old algebra is contained in the new, it is also possible to define an algebra $Q^\infty(\mathcal{A})$ generated by an infinite number of application of the ladder operator which then contains all the lower ones. More precisely, $Q^\infty(\mathcal{A})$ is the universal algebra generated from the algebras $Q^n(\mathcal{A})$ $n = 0,..., \infty$, modulo the identification: $X = Up(X)$

for all elements *X* of the algebra. The algebra $\mathcal{B} = Q^\infty(\mathcal{A})$ has the property that applying the ladder operator generates a new one which is isomorphic to the original.

$$Q(\mathcal{B}) \simeq \mathcal{B}$$

This raises an interesting question. Starting from a given algebra, is it possible that after only a finite number of applications of the ladder operator, you always arrive at the most complete algebra? Further steps may just create algebras isomorphic to the previous one. This is certainly the case for the algebras $\mathcal{B}$ which can be generated as above, but in the general case it is an open question.

## Multiple Quantisation

I would like to propose an interpretation of what the above construction means. The network of connections which appear in the construction of the ladder operator *Q* could be interpreted as Feynman diagrams. In that case *Q* can be interpreted as the process of quantisation. Here quantisation does not just mean a deformation in the sense that a quantum group is a deformation of a Lie group. It means the process of deriving a quantum theory from a classical one as outlined by Dirac using canonical quantisation or by Feynman using path integrals.

The application of *Q* many times is then multiple quantisation and suggests a connection with the long term research into multiple quantisation and ur-theory as studied by von Weizsäcker and his collaborators [6,7,8]. Ur-theory starts with bits of information which are quantised to give the group *SU(2)*. This group should be further quantised multiple times to construct a unified theory of physics.

If this interpretation is correct it also suggests a link between quantisation and dimension. The ladder operator *Q* produces *p*-brane structures of one higher dimension each time it is applied. It is also quite natural to think of quantisation as an operation which generates an extra dimension. Although 4-dimensional classical dynamics is only an approximation to the real 4-dimensional quantum physics, the 3-dimensional kinematic classical state is still preserved in the full quantum theory as the basis of the Hilbert space of states. In quantum field theory we do not usually think of the time dimension as being generated by quantisation. It is just the dynamics of the fields in space-time which are generated. However, in quantum gravity where the structure of space-time is itself part of the dynamics it *is* natural to regard time as being generated by quantisation and since the spatial dimensions are to be treated the same as the time dimension according to relativity, it is also natural to look to multiple quantisation as a mechanism for constructing the dynamics of space-time from more basic foundations.

String theorists have become expert at forming lower dimensional theories from higher dimensional ones by compactification of some of the dimensions. Their difficulty is that they do not have a rigorous foundation for the higher dimensional superstring and *p*-brane theories they begin with. I suggest that quantisation is the operation that can take string theories back up the dimensional ladder and that unlimited multiple quantisation is the way to understanding classical/quantum duality [9].

## Getting Knotted

The above superstring symmetry construction is all very well except that strings are not made from discrete fermionic partons. They are defined as continuous loops, but at the same time they may be topological objects which can be determined by discrete points. To try to capture this algebraically it may be necessary to envisage a string as being made from discrete partons with fractional statistics like anyons. Such partons may be repeatedly subdivided into partons with smaller fractional statistics until a continuous limit is found. If strings are truly topological, an infinite sequence of subdivisions may not be necessary.

When fractional statistics are introduced the links will have to be replaced by knots, and the supersymmetry algebras will need to be quantised. Now we *are* using the word quantisation in the sense of the deformation which is used to construct quantum groups but this may be related to multiple quantisation in a way that is not yet clear.

The use of braided structures will also resolve other problems which are inherent in the use of event-symmetric space-time. If the symmetric group acting on space-time events were part of universal symmetry then it is hard to see how parity could not be conserved since a mirror reflection of space-time is just a permutation of events. Furthermore, event-symmetry could be used to unravel topological solitons which are so important in string theories but which depend on the topology of space-time. These difficulties might be resolved if the symmetric group is replaced with the braid group acting on space-time events, especially if these events are tied together with strings which cannot pass through each other.

The generalisation of the quantisation operator $Q$ to strings of braided partons is not straight forward and I will only speculate on the way to proceed. To know how to do it correctly it is probably necessary to understand the construction in more basic algebraic terms than the combinatorial form in which it is presented above. If we replace the Lie superalgebras with semi-simple bi-algebras such as quantum groups then there is a natural algebraic interpretation of what the string networks are. They represent homomorphisms between tensor products of the bi-algebras.

A full construction will most likely be described in the language of category theory. Since there is thought to be a relation between quantum theories in n-dimensional space-time and the structure of n-dimensional categories, it is natural to look for a constructive operator which takes us from n-categories to (n+1)-categories and which can be interpreted as an algebraic form of quantisation. The result of multiple quantisation would then be an $\Omega$-category. I hope that further progress in this direction can be made in future work.

## Acknowledgement

I thank David Finkelstein for drawing my attention to the work on multiple quantisation by von Weizsäcker and for providing other insights.

# References

[1] E. Witten 1995, *String Theory Dynamics in Various Dimensions*, hep-th/9503124, Nucl. Phys. **B443**, 85-126

[2] P.K. Townsend 1995, *P-Brane Democracy*, hep-th/9507048

[3] O. Bergman 1996, *Evidence for String Substructure*, hep-th/9606039

[4] C.B. Thorn 1996, *Substructure of String*, hep-th/9607204

[5] P.E. Gibbs 1996, *The Principle of Event-Symmetry*, PEG-07-95, Int. J. Theor. Phys. **35**, 1037

[6] C.F. von Weizsäcker 1955, *Komplementarität und Logik*, Die Naturwissenschaften **42**, 521-529, 545-555

[7] C.F. von Weizsäcker 1958, *Die Quantentheorie der Einfachen Alternative*, Zeitschrift für Naturforschung, **13a**, 245-253

[8] T. Görnitz, D. Graudenz and C.F. von Weizsäcker 1992, *Quantum Field Theory of Binary Alternatives*, Int. J. Theor. Phys., **31**, 1929-1960

[9] M.J. Duff 1994, *Classical Quantum Duality*, hep-th/9410210